# Evaluating Dynamic Linking through the Query Process using the Licas Test Platform

Kieran Greer, Distributed Computing Systems, Belfast, UK.
http://distributedcomputingsystems.co.uk
Version 1.2

*Abstract:* A novel linking mechanism has been described previously [4] that can be used to autonomously link sources that provide related answers to queries executed over an information network. The test query platform has now been re-written resulting in essentially a new test platform using the same basic query mechanism, but with a slightly different algorithm. This paper describes recent test results on the same query test process that supports the original findings and also shows the effectiveness of the linking mechanism in a new set of test scenarios.

## 1  Introduction

This paper revisits the tests performed using the linking mechanism of the licas system [4][5]. The linking mechanism can be used to dynamically link sources, or nodes in a network, depending on arbitrary values as indicated by the user. In this case, source information nodes are linked based on the values entered as part of a query process. A query system has been written on-top of the licas system that can generate random networks and queries and store statistics as to how well the system optimises the query performance. Optimisation is measured in terms of how much the system reduces the search process by, but it must also consider any relative loss in the quality of answer (or QoS) resulting in the reduction of the search space. The query system has been re-written since the first set of tests, but uses the same general query process and so these tests can be compared directly with the previously published results. For this set of tests, the additional optimising features have been removed (borrow memory or local view) and so these tests measure more accurately the effectiveness of the linking mechanism by itself. There are also some other more minor changes, but results show that the linking mechanism is still as effective as for the previous query process.

The rest of the paper is organised as follows: section 2 gives a review of the licas system and linking mechanism. Section 3 describes some related work in the area of querying and linking distributed information sources. Section 4 describes the new query test platform, while section 5 gives the results of the recent sets of tests. Finally, section 6 gives some conclusions to the work.

## 2    Licas System / Source Linking Review

The licas system is a framework for building autonomous service-based networks. The framework provides a server on which to load services, a communication mechanism and a linking mechanism. An open source version of this system can be found at sourceforge.net[1]. As it is a framework, it is very lightweight, but requires the programmer to add most of the service-specific implementation him/herself. One feature that is built into the framework however is a dynamic linking mechanism that allows the services, or loaded components, to self-organise, by creating and then removing dynamic links between themselves. The linking mechanism contains three levels, where references to source nodes are added at the bottom level and must then be moved up to the top level before they are returned as reliable links. Source references are moved though reinforcing a weight value, where if their weight passes a particular threshold they are moved up to the next level, but the weight can also be decremented, when they are then moved down a level again. There is also a memory restriction at each level, allowing only a certain number of references to be stored. If the level is full, then a new reference can be stored there only after an existing one has been removed. The linking mechanism is generic and so it is possible to link any sort of data. The effectiveness of the links is determined only by the quality of the information that is used to create them. Two sources of information are linked through a path of concepts that describes how they are related to each other, where one application of the linking mechanism is to optimise the network for querying. If a query is executed on the network, then there might be many potential sources that can answer the query. If these numbers of sources can be reduced, then the query process will be much more efficient.

The path that is used to describe the link is critical and must contain the correct concept values to make it sufficiently accurate. For example, if the path was to link concepts through

---

[1] http://licas.sourceforge.net

evaluating a query part such as: 'A.Value1 GT B.Value2', then the path would need to contain all of the relevant information and would look like:

*B source instance – Value2 type – A source type – Value1 type – GT operator –> References to A source instances*.

This would mean that if the system then looked for a 'B' source type with a 'Value2' value larger than an 'A' source type's 'Value1' value, the path described would reference the appropriate 'A' source instances from a particular 'B' source instance.

A typical scenario for this is the Semantic Web [2], where a query would ask to retrieve information from more than one Web resource. This is different to a direct information retrieval request, asking for Web pages that contain particular keywords, for example. The Semantic Web adds meaning to the requests and allows the query system to perform some level of reasoning. It is similar to executing a database query over several relational tables of information, instead of retrieving information from a single table of information. Work however is now being carried out to make structured data records available and queryable over the Internet, which has resulted in the concept of a Data Web. Note the difference between this and the current Web that is essentially Web pages composed of completely unstructured natural language text. This is also a first step on the path towards the full Semantic Web, but will make structured data only available. The full Semantic Web will widen the scope such that both structured data and even what is traditionally thought of as unstructured or semistructured content will also be available.

With the Internet however, there is also the fact that potentially many sources can answer the query. For example, any number of Web pages can give you information about 'hit music records' and a querying mechanism cannot query all of them. If however, you ask about 'hit music records' and 'your favourite group', then if there are linked sources with this information, the querying mechanism will be much more efficient. It also has by its very nature, the potential to simply improve the quality of information, by indicating what sources are a good match for each other.

## 3   Related Work

There are now a number of Web browsers that focus specifically on querying the Semantic Web, for example [1] or [9]. These both show and note the importance of dynamic linking and the mechanism for building up a reliable structure to query over. In [1] they note that while a Web browser navigates along links between documents, a Semantic Web browser navigates along relationships (predicates) in a Web of concepts. This is more like an ontology construction. In [9] they note that the links are more important for personalisation, where the Web browser learns your preferences and can decide what pages to link, through intended meaning. This is now important because of the vast amount of information that is available. Also, because not all of the related pages are directly linked through hyperlinks, much time could be spent searching for and comparing them; where personalisation will greatly speed this process up. Existing Semantic Web search engines include Swoogle[2] and SHOE[3].

The paper [10] gives some nice examples and reasons why it is important to be able to query over the Semantic Web [1]. RDF [8] appears to be the preferred ontology or data model for the Semantic Web and in [10] they argue that the relational model is no longer appropriate for some of the information that is stored on the Internet, such as for life sciences or lifelog management. The relational model is largely a static model, but in the area of lifelogs with pictures, videos, music, conversation records, etc., any relationship can be made between any two lifelogs. These relationships can also change as the lifestyle or job of a person changes, creating new kinds of lifelogs and relationships. In this case, a static-table approach of the relational data model is not appropriate, but the relationship model of something like RDF would be.

XML-based query languages tend to express the query in terms of a pattern of concepts that needs to be matched. This is because the XML data has depth or structure to it, with sub-concepts of other concepts, and so a pattern can be required to avoid concept mismatch. This pattern can be a single path description or a more complex schema that needs to be matched to. In [10] they use a visual aid (SQBE) to map a SPARQL query to the underlying RDF data model that would be queried on the Semantic Web. This also shows that it is now important to be able to link different sites of information together in a single query. In [7] they describe another query language called RDF-QEL-i that can also be used as part of a visual graph, or pattern-based, query construction process. These languages originated with, and can still have, an SQL-like structure, but possibly with additional constructs, such as 'typeof', or 'subclassof'. One example of this is the RQL language [3]. In [6] they describe work that tries to improve the efficiency of information retrieval using these sorts of query languages. While there may be pattern matching, the basic 'select-from-where' structure is still required. This is because you need to specify what to request, what sources to query and also the comparison rules or conditions, under which the information from those sources should be retrieved. This means that with a much simpler data model, a basic select-from-where query can be used to show the effectiveness of a query process linking related information sources, which is what is being described in this paper.

---

[2] http://swoogle.umbc.edu/
[3] http://www.cs.umd.edu/projects/plus/SHOE/search/

## 4  The New Query Test Platform

The linking mechanism has already been extensively tested and the results presented in [4]. Although the testing was extensive (many test runs over different test scenarios), the problem is too large to give any really definitive conclusions. It would need to be tried out for real to determine exactly how it might work. The querying system has now been re-written since the previous results were released. This was done to try and improve the code, where there were some minor restrictions on the previous version, and also to make the code independent of the previous project. This has also led to a slightly different querying system that uses the same query process, meaning that any test results would not necessarily need to be the same as the previous results, but could be directly compared to them. So these new tests, at least partially, provide results on the same mechanism from a different system. The previous tests also tried to give an overview of how the system might work with all of the potential features being used. As well as a global linking mechanism, this included a local view and some memory management. The local view in particular would significantly reduce the search space, possibly at the cost of some accuracy. However, these additional features also obscured the true performance of the linking mechanism by itself and so this set of tests only demonstrates how well the linking mechanism by itself might work. The next section describes how the querying mechanism processes a query to try and answer it.

## 4.1 Example of Processing a Query Part

Random queries can be generated, with source types being allocated the names of letters and value types being allocated an additional numerical value to tell them apart. The source and value types can then be placed into bands and when one is required for the query, it is selected from one of the bands with a frequency that is dependent on a specified probability. So if there is a band with three source types and a probability value of 70%, then when the query requires a source type, it will select one of these three types 70% of the time. The queries are constructed with a maximum number of sources in the 'Select' and 'From' clauses specified as part of the test. The 'Where' clauses then use the 'From' clause source types to randomly construct comparison constraints over those values. These comparisons can be of any type, but in these tests only the equivalence comparison is used. These tests could then also be considered to be matching symbolic concepts instead. The query variability then comes from what source and value type combination to compare with each other. Following is an example query:

*Select A.Value1, B.Value2 From A, B, C Where (A.Value3 EQ B.Value4) And (B.Value2 EQ C.Value1 Or B.Value3 EQ C.Value2)*

As described in [4], this statement contains information about source types, value types and comparison operators that are related to each other through the query. It is possible to use this information to create linking structures that describe different parts of the query. These structures can be made from sets of nested hashtables, for example, where the keys represent the path to the related source, which can be constructed from the source types, value types and operators that are used in the query (section 2). A query engine might typically perform the following steps to try and evaluate the first part of this query:

1. The first part of the query process is to evaluate the comparison 'B.Value3 EQ C.Value2'. If we evaluate a source with a specific value, for example 'B.Value3 EQ 10', then the comparison can be evaluated directly at the source. However, for these types of queries we are comparing two different source types, where there may be many instances of each type in the network. This means that we cannot know the optimal sources to use before the search and so all relevant sources need to be retrieved and compared to determine which combinations satisfy the comparison. It is this comparison that would link different web pages together, for example.

2. Sources are retrieved in the following order: if sources exist from a previous evaluation then these are used. If there has not been a previous evaluation with the specified source, then links can be used. If there are no links or previous evaluations, then a full search is performed.
3. The same process is performed for the comparison 'B.Value2 EQ C.Value1' and in fact with the 'OR' logical operator, the valid B and C source instances for the first two comparisons are then combined. It is then only the combined B source instances that can then be used for the other query evaluations.
4. This process filters through the whole query to evaluate each query part, with only the sources allowed from previous evaluations.

Some other query conditions now include the following:
- The query process can actually store links in either a forwards or a backwards direction. For example, evaluating (A.Value2 GT B.Value2) can store links from A to B, or from B to A, and also retrieve the links in the same direction. The test results of the following sections include results from both types of linking process, but in the main, used the forwards linking approach.
- Consider a query that asks for A.Value1 with a condition of (A.Value2 GT A.Value3). The search could return an 'A' source 'A3' with a 'Value3' value of '3' as part of the solution, but this source could have a very small 'Value1' value, meaning that it should not be part of the optimal solution set, even if it is part of the solution comparison. This also means that links cannot be stored from a specific source instance to itself, as the link can then be retrieved and used on either side of a comparison, when it might not be appropriate. So this evaluation would never include the same source type on both sides of the equation.
- If the query asks for two values from a particular source type, they should probably be retrieved from the same single source instance. For example, if a query is something like: *select A.Value4, A.Value2 from A where (A.Value4 EQ A.Value3)*, then the 'Value4' and 'Value2' values should probably be retrieved from the same source instance.

## 4.2 Linking Path Construction

It has been argued that adding a path of descriptive concepts to the links provides sufficient accuracy for the linking mechanism to be reliable. It was then argued that if the linking mechanism is reliable it can be used as part of some autonomous process, for constructing higher level, or more complex concepts, by linking or combining less complex ones. The problem however possibly transfers from how to make the link accurate enough, to how to use the correct concepts in the path, to make the link accurate enough. Other tests have shown that it is relatively easy to combine concepts presented as individual entities, so long as they are presented in the same order each time. For example, if a system is presented with the following fragments of information:

*ABC, BCD, GH, DEF, HIJ, CDEF, BC*

It can quickly determine that ABCDEF, for example, is a higher-level concept, by combining some of the fragments. Linking these individual concepts can be done by using a default path or no path information at all. The Internet-related source-value query mechanism of this section however requires the path to include the value and source types, and also the comparison operators. The query mechanism could also be considered to be an optimisation problem, where the 'Where' conditions are a set of constraints that need to be satisfied. It is then these constraints that need to be included in the path description to make the linking structure sufficiently accurate.

If the system is presented with only concept names and no constraint conditions, it can form the links without any real path structure. Adding noise, fuzziness, or some other quality measure might change this, but this does suggest however that the constraints on the linking process would be a good guide as to what to add to any path description. They restrict any possible answer set, because their conditions need to be met, and so they essentially define what additional accuracy is required. So in this sort of scenario, the query process would be critical in helping the network to actually structure itself. The information to be retrieved will need to satisfy any constraints in the query and therefore the link structure could be based on these constraint types or values. An autonomous system still needs to be able to recognise what a constraint is by itself, which might be another translation of the problem into another form. The queries therefore would be able to tell the system what variables would make sensible paths and also the exact value types to use, where a formal query language can be parsed. This is probably not a complete solution however, if information can be presented in any arbitrary way, but for a structured format, it could be possible as part of an autonomous system.

## 5   Tests and Results

There are some differences to these sets of tests and the previous tests. One critical factor that the previous sets of tests did not measure was the time factor. To measure this, these tests are run using a server. Any query must use the communication mechanism and invoke methods on sources running on a server, instead of querying everything locally. This is to try and mimic the time spent executing queries over the Internet. It turns out that the communication mechanism used up most of the querying time and so the time value saved was almost equal to the reduction in the number of nodes searched. There was a fear that manipulating links in the linking mechanism would be time costly, but the communication has proved to be much more time consuming. This might change however if sources are allowed to borrow memory off each other, but this is an additional feature and is not compulsory.

These test results largely copy the previous process of executing random queries on networks of different sizes to try and determine what sort of variation the linking mechanism can cope with. These tests only consider queries with the equivalence operator, that is, each condition is an equivalence comparison. They also always add the full search counts to the linked search stats when the linked search does not return an answer. These tests also only use the global linking structure and do not use a local view. They do however still limit the memory to 50 entries for each level of the linking structure. These tests verify some of the conclusions or assumptions of the previous tests and so in the main support the previously published results. The test process is explained in detail in [4], which is the same process used here, the only difference being that in these tests, values were measured at intervals of 2500 queries over a total of 40000 queries per test. The previous tests measured values after every 5000 queries. The other difference is that these tests in general used the forwards linking direction, but comparisons with the backwards linking direction are given in the conclusions section.

## 5.1 70:30 split on Networks of Varying Sizes

In these tests, the queries were split in a 70:30 manner, with the network size also varying. Three network sizes were tested – 5 source types with 10 instances of each, 10 source types with 15 instances of each and 10 source types with 30 instances of each. The 70:30 split would mean that 70% of the time a query would use a source type from one of 2 (5 types) or 3 (10 types) and 30% of the time from one of 3 (5 types) or 7 (10 types) when formulating a query. There were always 5 value types, with values in the range 1 to 10, and a split of 70% for 2 types and 30% for 3 types. Graph 1 shows the comparative reduction in the number of nodes searched, while Graph 2 shows the comparative loss in the quality of the answer compared to an optimal answer.

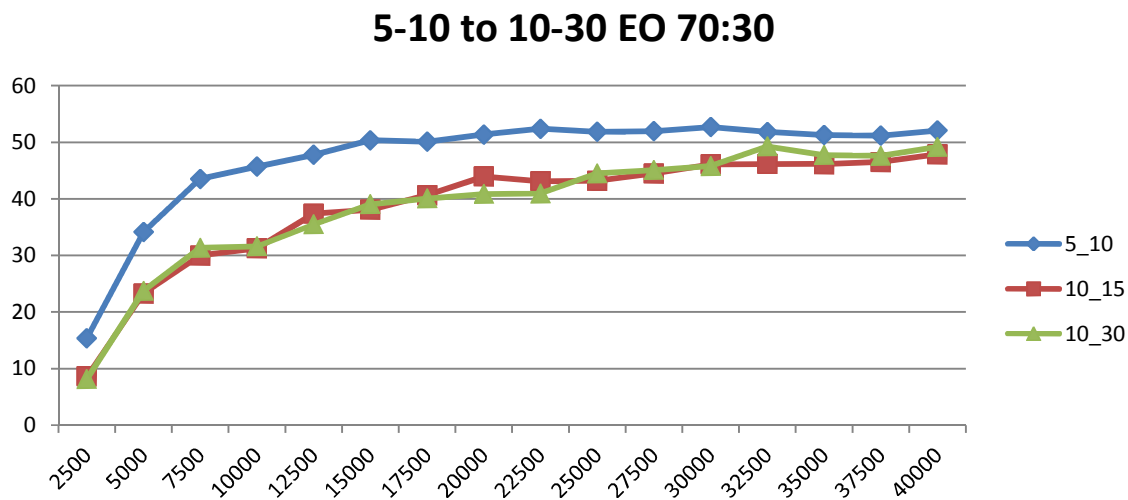

Graph 1. Comparison of the amount of search reduction for three different network sizes for equivalence only queries generated from a 70:30 split. Either 5 or 10 source types and 10, 15 or 30 instances of each.

These results show again that the quality of answer can be improved together with a search reduction in the number of nodes searched. For example, the transition for the 10-30 network between 17500 queries and 20000 queries. Quite a lot was made previously of the fact that the system can peak and then start to perform worse with the addition of more links. This led to the idea of the system autonomously monitoring itself because it could tell when it had reached its optimum level. This is maybe not quite as pronounced in these tests, but the test values clearly oscillate from a larger to a smaller value, indicating a peak level, then a lower value and then an improvement again, etc. So it is still the case that if the system could recognise this it could tell for itself when performance was dropping or when a general peak level had been reached. In these tests however, the fact that the values seem to level off would also be a good indicator.

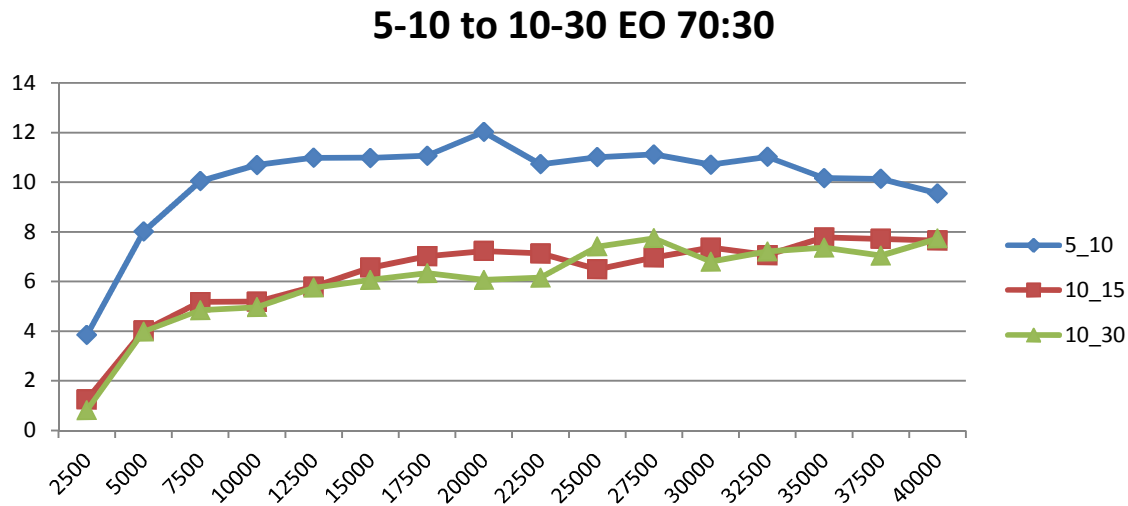

Graph 2. Comparison of the loss in QoS for three different network sizes for equivalence only queries generated from a 70:30 split. Either 5 or 10 source types and 10, 15 or 30 instances of each.

### 5.2 10-30 Size Network with 70:30 or 90:10 Skewed Splits

In these tests, the network size was kept the same with 10 source types and 30 instances of each, but the queries were split in either a 70:30 or a 90:10 manner. The 90:10 split placed the same number of source or value types in each group, only the frequency of selecting from that group changed to 90% of the time for the first group and 10% of the time for the second group. This would mean that there was less variability with the 90:10 queries and so the linking mechanism should work better because it has to deal with a smaller variation in what it needs to link. Graph 3 shows the comparative reduction in the number of nodes searched, while Graph 4 shows the comparative loss in the quality of the answer compared to an optimal answer. For the 90:10 split the quality of answer actually starts much worse and gradually improves, which is the opposite for the 70:30 split. With a 90:10 split, any links that are added are more likely to be used and so if they are not the best links this will lead to a worse answer. As the links build up and become more accurate, the QoS appears to improve. Adding more links in this case, will mean more chance of returning a link with an optimal value. With the more variable 70:30 queries, the initial links might not be used to answer all parts of a query. For the stats however, if a link is ever used then the query is considered to be answered by a linked search. As the link numbers build up, they will then be used in more parts of a query answer and if they do not contain optimal values, they could reduce the total value compared to an optimal result. This can be confirmed by looking at how many times either method returns a result that is the same as the optimal result and how many times it returns a worse result. The 70:30 split starts off answering queries and returning an optimal value possibly 2400 times out of 2500 queries, which then reduces to closer to 1800 times out of 2500 queries. With a 90:10 split, the optimal value is initially achieved possibly 1850 times for 2500 queries, but this then increases to closer to 1900 times for 2500 queries.

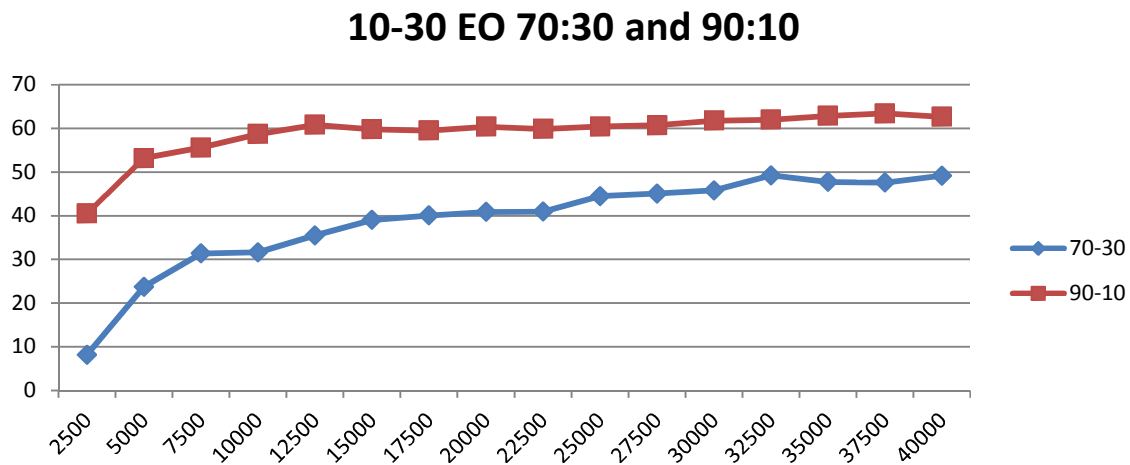

Graph 3. Comparison of the amount of search reduction for equivalence only queries generated from either a 70:30 or a 90:10 split. 10 source types and 30 instances of each.

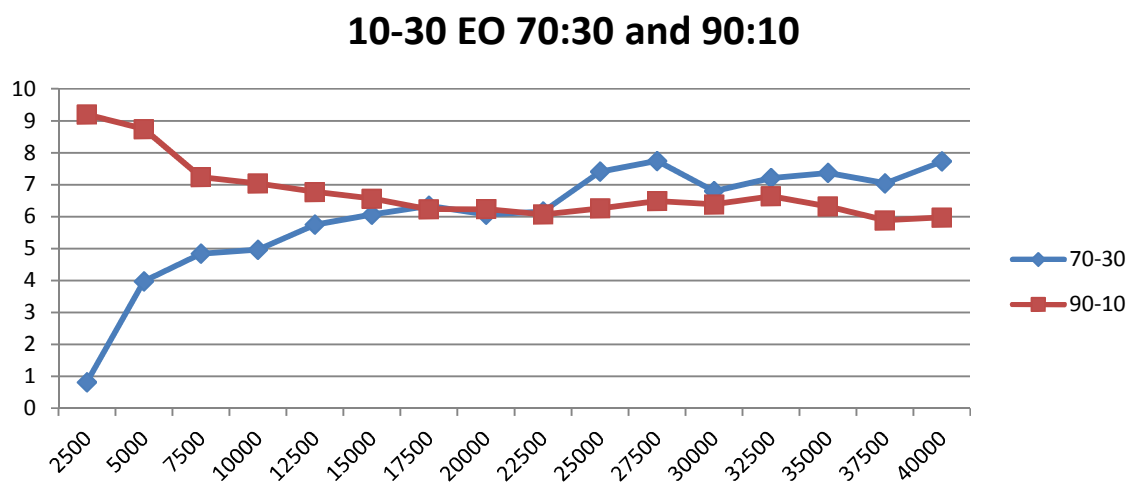

Graph 4. Comparison of the loss in QoS for equivalence only queries generated from either a 70:30 or a 90:10 split. 10 source types and 30 instances of each.

When an optimal result is not returned, results show that the result returned is generally quite poor and can easily be 20-30% worse than the optimal value. So with 90:10 split, because of less variability, relevant links are added almost immediately but then need to be made more accurate. With the 70:30 split, there are fewer links being used to answer queries initially, but as the numbers build up and they increasingly replace any full searches, the QoS starts to depreciate more. This depreciation appears to eventually level off however. Overall, the 90:10 split shows a better QoS than the 70:30 split and also a better search reduction.

# 6  Conclusions

While the results in these tests are slightly different, they verify the reliability of the linking process as a whole. Previous test results including a view [4] showed for similar configurations, a minimum loss in QoS of just over 4% and search reduction of around 85%. But the query process for those tests was slightly different. These results without a view actually show a slightly worse value of just below 6%, while a search reduction of between 50-65% is still reasonably healthy. New view tests also show that the current system does not produce the same stats as the previous system. Two different versions of the view have now been tested. One gives close to 80% search reduction with possibly around 12% loss in QoS, while the other gives a smaller search reduction that is closer to 60%, with a QoS loss of just over 6%. The 80% search reduction results from a forwards type of query linking process, while the 60% results from the intuitively more correct backwards linking (see section 4.1). The improvement in the search however is also a result of the fact that the forwards linking process answers more queries and therefore does not add as many of the full search results to its total. If the full search results are not taken into account, then both processes have a similar search reduction, but the backwards linking process is then more accurate. In that case, a direct comparison of the searches for the linked and full search versions for either method can give a search reduction close to 95%. That is - only compare the linked search stats by themselves with the full search stats, over the same number of queries. In general however, the forwards linking method would produce a better search reduction, but with a worse QoS.

The new system however is able to produce even more random queries, so the differences in the results are also because of the more random data, including queries that compare the same source types with each other. The previous tests did not have any of these types of query. One other result was for a 90:10 split without a view and the backwards linking direction. This did not reduce the search by very much, but equivalently, reduced the QoS by a minimal amount as well. If the full search counts were removed, then a real reduction in the search would be obtained. The main conclusion is probably the fact that these tests back up the previous tests in confirming the effectiveness of the linking mechanism, although certain results also give some insights into how the links are specifically used as well.